\documentclass[10pt,conference,comsocconf]{IEEEtran}

\usepackage[acronym]{glossaries}
\usepackage[T1]{fontenc}

\usepackage{blindtext}
\usepackage{graphicx}
\usepackage{subcaption}
\usepackage{placeins} 

\usepackage{multirow} 
\usepackage{tikz} 
\usetikzlibrary{calc,fit,shapes}
\usetikzlibrary{hobby,backgrounds,trees}
\usepgflibrary{arrows}
\usepackage{varwidth}
\usetikzlibrary{positioning,arrows}

\usepackage{url}     
\usepackage[pdftex, hidelinks]{hyperref}
\hypersetup{
    bookmarks=false,
}

\usepackage{fancyhdr}
\usepackage{tcolorbox}

\usepackage{multirow}

\begin{document}

\thispagestyle{empty} 

\vfill 

\begin{center}
\centering
\begin{tcolorbox}[
    colback=gray!10, colframe=black, fonttitle=\bfseries,
    width=0.9\textwidth, boxrule=1pt, arc=5pt, outer arc=5pt,
    boxsep=10pt, left=10pt, right=10pt, top=10pt, bottom=10pt
]
\centering
\textbf{THIS IS AN AUTHOR-CREATED POSTPRINT VERSION.}

\vspace{0.3cm}

\textbf{Disclaimer:}  
This work has been accepted for publication in the \textit{Joint European Conference on Networks and Communications \& 6G Summit (EuCNC/6G Summit)}, 2025.  

\vspace{0.3cm}

\textbf{Copyright:}  
© 2025 IEEE. Personal use of this material is permitted. Permission from IEEE must be obtained for all other uses, in any current or future media, including reprinting/republishing this material for advertising or promotional purposes, creating new collective works, for resale or redistribution to servers or lists, or reuse of any copyrighted component of this work in other works.

\vspace{0.3cm}


\end{tcolorbox}
\end{center}

\vfill 

\clearpage
\setcounter{page}{1}

\newacronym{3GPP}{3GPP}{3rd Generation Partnership Project}

\newacronym{5G}{5G}{5th Generation}
\newacronym{5G-ACIA}{5G-ACIA}{5G Alliance for Connected Industries and Automation}
\newacronym{5GC}{5GC}{5G Core}
\newacronym{5QI}{5QI}{5G QoS Identifier}

\newacronym{AF}{AF}{Application Function}
\newacronym{AMBR}{AMBR}{Aggregate Maximum Bit Rate}
\newacronym{ARP}{ARP}{Allocation and Retention Priority}

\newacronym{BLER}{BLER}{Block Error Rate}
\newacronym{BSR}{BSR}{Buffer Status Report} 

\newacronym{C2C}{C2C}{Controller-to-controller}
\newacronym{C2D}{C2D}{Controller-to-device}
\newacronym{CCDF}{CCDF}{Complementary Cumulative Distribution Function}
\newacronym{CDF}{CDF}{Cumulative Distribution Function}
\newacronym{CG}{CG}{Configured Grant}
\newacronym{CNC}{CNC}{Centralized Network Configuration}
\newacronym{CQI}{CQI}{Channel Quality Indicator}

\newacronym{D2Cmp}{D2Cmp}{Device-to-computer}
\newacronym{DCI}{DCI}{Downlink Control Information}
\newacronym{DG}{DG}{Dynamic Grant}
\newacronym{DL}{DL}{Downlink}
\newacronym{DN}{DN}{Data Network}
\newacronym{DRB}{DRB}{Data Radio Bearer}
\newacronym{DRL}{DRL}{Deep Reinforcement Learning}
\newacronym{DSCP}{DSCP}{Differentiated Services Code Point}
\newacronym{DS-TT}{DS-TT}{Device-side Translator}
\newacronym{D-TDD}{D-TDD}{Dynamic TDD}

\newacronym{E2E}{E2E}{End-to-end}
\newacronym{eMBB}{eMBB}{enhanced Mobile Broadband}

\newacronym{FDD}{FDD}{Frequency Division Duplex}

\newacronym{gNB}{gNB}{next generation Node B}
\newacronym{gPTP}{gPTP}{generic Precision Time Protocol}
\newacronym{GTP}{GTP}{GPRS Tunneling Protocol}

\newacronym{HARQ}{HARQ}{Hybrid Automatic Repeat Request}
\newacronym{HP}{HP}{High Priority}
\newacronym{HMP}{HMP}{High-Medium Priority}


\newacronym{L2C}{L2C}{Line controller-to-controller}
\newacronym{LMP}{LMP}{Low-Medium Priority}
\newacronym{LP}{LP}{Low Priority}
\newacronym{LTE}{LTE}{Long Term Evolution}

\newacronym{MAC}{MAC}{Medium Access Control}
\newacronym{MBS}{MBS}{Multicast–Broadcast Services}
\newacronym{MBSFN}{MBSFN}{Multicast-Broadcast Single-Frequency Network}
\newacronym{MB-UPF}{MB-UPF}{Multicast/Broadcast UPF}
\newacronym{MCS}{MCS}{Modulation and Coding Scheme}
\newacronym{MNO}{MNO}{Mobile Network Operator}
\newacronym{Multi-TRP}{Multi-TRP}{Multiple Transmission and Reception Point}

\newacronym{NIC}{NIC}{Network Interface Card}
\newacronym{NW-TT}{NW-TT}{Network-side Translator}
\newacronym{NR}{NR}{New Radio}
\newacronym{NPN}{NPN}{Non-Public Networks}
\newacronym{NTP}{NTP}{Network Time Protocol}

\newacronym{OFDMA}{OFDMA}{Orthogonal Frequency-Division Multiple Access}
\newacronym{OFDM}{OFDM}{Orthogonal Frequency-Division Multiplexing}
\newacronym{O-RAN}{O-RAN}{Open Radio Access Network}

\newacronym{PCP}{PCP}{Priority Code Point}
\newacronym{PDCCH}{PDCCH}{Physical Downlink Control Channel}
\newacronym{PDCP}{PDCP}{Packet Data Convergence Protocol}
\newacronym{PDR}{PDR}{Packet Detection Rule}
\newacronym{PDSCH}{PDSCH}{Physical Downlink Shared Channel}
\newacronym{PDU}{PDU}{Packet Data Unit}
\newacronym{PLC}{PLC}{Programmable Logic Controller}
\newacronym{PRACH}{PRACH}{Physical Random Access Channel}
\newacronym{PTP}{PTP}{Precision Time Protocol}
\newacronym{PUCCH}{PUCCH}{Physical Uplink Control Channel}
\newacronym{PUSCH}{PUSCH}{Physical Uplink Shared Channel}

\newacronym{QAM}{QAM}{Quadrature Amplitude Modulation}
\newacronym{QoE}{QoE}{Quality of Experience}
\newacronym{QoS}{QoS}{Quality of Service}
\newacronym{QFI}{QFI}{QoS Flow ID}

\newacronym{RAN}{RAN}{Radio Access Network}
\newacronym{RB}{RB}{Resource Block}
\newacronym{RLC}{RLC}{Radio Link Control}
\newacronym{RRC}{RRC}{Radio Resource Control}
\newacronym{RTC1}{RTC1}{Real Time Class 1}

\newacronym{SCS}{SCS}{Sub-Carrier Spacing}
\newacronym{SMF}{SMF}{Session Management Function}

\newacronym{SINR}{SINR}{Signal-to-Interference-plus-Noise Ratio}
\newacronym{SPS}{SPS}{Semi-Persistent Scheduling}

\newacronym{TAS}{TAS}{Time-Aware Shaping}
\newacronym{TDD}{TDD}{Time Division Duplex}
\newacronym{TSC}{TSC}{Time-Sensitive Communication}
\newacronym{TSN}{TSN}{Time-Sensitive Networking}
\newacronym{TTI}{TTI}{Transmission Time Interval}

\newacronym{UDP}{UDP}{User Datagram Protocol}
\newacronym{UE}{UE}{User Equipment}
\newacronym{UFTP}{UFTP}{UDP-based File Transfer Protocol}
\newacronym{UL}{UL}{Uplink}
\newacronym{UPF}{UPF}{User Plane Function}
\newacronym{uRLLC}{uRLLC}{Ultra-Reliable and Low-Latency Communications}

\newacronym{VLAN}{VLAN}{Virtual Local Area Network}
\newacronym{VNI}{VNI}{Virtual Network Identifier}
\newacronym{VTEP}{VTEP}{VxLAN Tunnel End Point}
\newacronym{VxLAN}{VxLAN}{Virtual Extensible LAN}

\pagestyle{fancy}
\fancyhf{} 
\renewcommand{\headrulewidth}{0pt}
\fancyhead{} 

%
\title{Empirical Analysis of 5G TDD Patterns Configurations for Industrial Automation Traffic}

\author{\IEEEauthorblockN{Oscar Adamuz-Hinojosa\IEEEauthorrefmark{1}, Felix Delgado-Ferro\IEEEauthorrefmark{1},  Núria Domènech\IEEEauthorrefmark{3}, Jorge Navarro-Ortiz\IEEEauthorrefmark{1}, Pablo Muñoz\IEEEauthorrefmark{1},\\ Seyed Mahdi Darroudi\IEEEauthorrefmark{3}, Pablo Ameigeiras\IEEEauthorrefmark{1}, Juan M. Lopez-Soler\IEEEauthorrefmark{1}}

\IEEEauthorblockA{\IEEEauthorrefmark{1}Department of Signal Theory, Telematics and Communications, University of Granada.}
\IEEEauthorblockA{\IEEEauthorrefmark{3}Neutroon Technologies S.L., Barcelona, Spain.\\ Email: \{oadamuz,felixdelgado,jorgenavarro,pabloml,pameigeiras,juanma\}@ugr.es\IEEEauthorrefmark{1}\IEEEauthorrefmark{2}  \\ \{nuria.domenech,mahdi.darroudi\}@neutroon.com\IEEEauthorrefmark{3}}
}

\markboth{Journal of \LaTeX\ Class Files,~Vol.~6, No.~1, January~2007}%
{Shell \MakeLowercase{\textit{et al.}}: Bare Demo of IEEEtran.cls for Journals}

\maketitle

\begin{abstract}
The digital transformation driven by Industry 4.0 relies on networks that support diverse traffic types with strict deterministic end-to-end latency and mobility requirements. To meet these requirements, future industrial automation networks will use time-sensitive networking, integrating 5G as wireless access points to connect production lines with time-sensitive networking bridges  and the enterprise edge cloud. However, achieving deterministic end-to-end latency remains a challenge, particularly due to the variable packet transmission delay introduced by the 5G system. While time-sensitive networking bridges  typically operate with latencies in the range of hundreds of microseconds, 5G systems may experience delays ranging from a few to several hundred milliseconds. This paper investigates the potential of configuring the 5G time division duplex pattern to minimize packet transmission delay in industrial environments. Through empirical measurements using a commercial 5G system, we evaluate different TDD configurations under varying traffic loads, packet sizes and full buffer status report activation. Based on our findings, we provide practical configuration recommendations for satisfying requirements in industrial automation, helping private network providers increase the adoption of 5G.

\end{abstract}

\begin{IEEEkeywords}
5G, Industry 4.0, TDD pattern, testbed
\end{IEEEkeywords}

\section{Introduction}
The integration of recent advancements in IoT, Cyber-Physical Systems, cloud computing, and Artificial Intelligence into industrial production is already transforming traditional industries, laying the groundwork for Industry 4.0~\cite{Wollschlaeger2017}. This requires network platforms that enable efficient, low-latency and reliable communication across all components. To address this, IEEE introduced \gls{TSN}, a set of standards that enhance industrial networks with synchronization, stream reservation, traffic shaping, scheduling, preemption, traffic classification, and seamless redundancy~\cite{RostSOTA}. However, wired industrial networks are limited in scalability and flexibility due to the complexity of adding or relocating equipment, restricting mobility and coverage to cabled areas.

To overcome these issues, researchers are exploring the integration of the \gls{5G} system as \gls{TSN} bridges  within \gls{TSN} networks~\cite{MahmoodSOTA}. In this setup, production lines in a factory connect wirelessly to the \gls{5G} system, which then interfaces with the enterprise edge cloud through \gls{TSN} bridges~\cite{5GACIA-whitepaperI}. This integration presents challenges due to the differences between \gls{TSN} and \gls{5G} technologies~\cite{Sasiain2024}. One key challenge is maintaining deterministic \gls{E2E} communications, which refers to ensuring the packet transmission delay remains consistently below a predefined threshold with very high reliability. This is difficult to achieve due to the non-deterministic latency and jitter caused by the time-varying wireless channel in the \gls{5G} system. While \gls{TSN} bridges typically have latency in the hundreds of microseconds~\cite{Seijo2022}, \gls{5G} systems can experience delays ranging from milliseconds to several hundred milliseconds~\cite{Sasiain2024}.

In this regard, the use of \gls{NPN} allows for customized \gls{5G} network configurations, which present the challenge of adapting network performance to the service needs of each vertical industry~\cite{Prados2021}. To minimize packet transmission delay in the \gls{5G} system, \gls{NPN} operators can customize the \gls{TDD}\footnote{\gls{5G} networks predominantly use \gls{TDD} over \gls{FDD} because it offers greater spectral efficiency, flexibility in asymmetric traffic management, and better resource utilization~\cite{GSMA2020_5GTDD}.} pattern, which involves controlling the allocation of transmission slots between \gls{UL} and \gls{DL}  and optimizing the frequency of switching between sending and receiving data~\cite{Kim2020}. The challenge in configuring \gls{TDD} is to meet strict latency requirements while considering factors such as the lack of real-world deployment studies, the range of possible parameter combinations and the strong dependence on network load.

\subsection{Related Works}
Recent research on \gls{TDD} has focused on dynamic configurations to optimize \gls{UL}/\gls{DL} time slot allocation using machine learning. For example, the authors of \cite{Jeong2024} use convolutional long short-term memory models combined with \gls{DRL} for real-time adjustments, while the authors of \cite{Bagaa2021, Boutiba2024} propose a \gls{DRL}-based approach to address challenges like buffer overflow and varying traffic demands. \gls{TDD} configurations have also been explored for network slicing~\cite{Shrivastava2018}, enabling service-specific resource allocation. However, these studies rely on simulations, which may overlook practical implementation challenges. For a broader overview of dynamic \gls{TDD} pattern configurations, we refer to the comprehensive survey in~\cite{Kim2020}.

In contrast, some works have implemented solutions in commercial base stations. For instance, the authors of \cite{Boutiba2023} integrate a \gls{DRL}-based solution into an \gls{O-RAN}-supported base station, and other authors in \cite{Lai2023} investigate the trade-off between data rate and latency in an Open Air Interface-based base station. Despite these efforts, none of these studies address the specific needs of industrial environments, where traffic demands impose strict requirements on packet transmission delay and reliability.

In an industrial context, the authors of \cite{Esswie2021} propose a service-aware \gls{TDD} framework that balances latency and data rate through dynamic scheduling and reinforcement learning. While the study provides valuable insights, it is based on simulations and does not consider real-world aspects.

\subsection{Contributions}
This paper investigates the impact of different \gls{TDD} patterns on \gls{5G} systems, based on empirical measurements from a commercial \gls{5G} system in a laboratory testbed. We assess packet transmission delay under varying conditions, including traffic load, packet size, and the effect of full \gls{BSR} activation at the \gls{UE}. The evaluation covers both \gls{UL} and \gls{DL} transmissions. Based on the results, we offer configuration recommendations for private industrial network operators to optimize support for diverse industrial automation traffic types. Our findings contribute to the integration of \gls{5G} into \gls{TSN}-based industrial networks, paving the way for deterministic communications.

The paper is structured as follows. Section~\ref{sec:TDDpatterns} reviews industrial traffic types and their requirements; and the concepts of \gls{5G} \gls{TDD} pattern and \gls{UE} \gls{BSR}. In Section~\ref{sec:ProofConceptandResults}, we describe our proof of concept and discuss the obtained results. We also provide recommendations for configuring \gls{TDD} patterns in industrial environments. Finally, Section~\ref{sec:conclusions} summarizes key conclusions and outlines potential areas for future work.

\section{Industrial Automation Traffic and 5G Features under Evaluation}\label{sec:TDDpatterns}

\subsection{Traffic Types for Industrial Automation}\label{sec:BackgroundIndustrialNetworks}
Industrial networks involve the following traffic types as described in Table \ref{tab:TrafficTypes}~\cite{5GACIA-whitepaperI}:
\begin{itemize}
    \item \textit{Network Control}: manages tasks like time synchronization, network redundancy, and topology detection.
    \item \textit{Cyclic Synchronous}: coordinates regular synchronized user plane data exchanges between devices.
    \item \textit{Cyclic Asynchronous}: involves periodic but unsynchronized user plane data exchanges.
    \item \textit{Events}: trigger messages based on metric changes.
    \item  \textit{Mobile Robots}: include movement control, task assignment, and sensor data. 
    \item \textit{Augmented reality}: provides real-time video and maintenance instructions.
    \item  \textit{Configuration and diagnostic}: handle non-critical data like device configuration and firmware downloads.
\end{itemize}

\begin{table}[b!]
\centering
\caption{Traffic Types and Performance Requirement}
\label{tab:TrafficTypes}
\begin{tabular}{|c|c|c|c|}
\hline
\multirow{2}{*}{\textbf{\begin{tabular}[c]{@{}c@{}}Traffic \\ Types \cite{5GACIA-whitepaperI}\end{tabular}}} & \multirow{2}{*}{\textbf{\begin{tabular}[c]{@{}c@{}}Periodic /\\ Sporadic\end{tabular}}} & \multirow{2}{*}{\textbf{\begin{tabular}[c]{@{}c@{}}E2E Delay \\ bound (ms)\end{tabular}}} & \multirow{2}{*}{\textbf{\begin{tabular}[c]{@{}c@{}}Typical Data \\ Size (Byte)\end{tabular}}} \\
                                                                                                                              &                                                                                                                                                                                                                                &                                                                                           &                                                                                               \\ \hline
\begin{tabular}[c]{@{}c@{}}\textbf{Network}\\ \textbf{Control}\end{tabular}                                                                     &  Periodic                                                                                & $[50, 1000]$                                                                    & \begin{tabular}[c]{@{}c@{}}Variable \\ $[50, 500]$\end{tabular}                     \\ \hline
\textbf{Isochronous}                                                                                                          &  Periodic                                                                                & $[0.1, 2]$                                                                    & \begin{tabular}[c]{@{}c@{}}Fixed\\ $[30, 100]$\end{tabular}                         \\ \hline
\textbf{\begin{tabular}[c]{@{}c@{}}Cyclic\\ Synchronous\end{tabular}}                                                         &  Periodic                                                                                & $[0.5, 1]$                                                                     & \begin{tabular}[c]{@{}c@{}}Fixed \\ $[50, 1000]$\end{tabular}                       \\ \hline
\textbf{\begin{tabular}[c]{@{}c@{}}Cyclic \\ Asyncrhonous\end{tabular}}                                                       &  Periodic                                                                                & $[2, 20]$                                                                     & \begin{tabular}[c]{@{}c@{}}Fixed \\ $[50, 1000]$\end{tabular}                       \\ \hline
\textbf{Events}                                                                                                               &  Sporadic                                                                                & $[10, 2000]$                                                                   & \begin{tabular}[c]{@{}c@{}}Variable \\ $[100, 1500]$\end{tabular}                   \\ \hline
\begin{tabular}[c]{@{}c@{}} \textbf{Mobile}\\ \textbf{Robots} \end{tabular}                                                                       &  Both                                                                                    & $[1, 500]$                                                                      & \begin{tabular}[c]{@{}c@{}}Variable \\ $[64, 1500]$\end{tabular}                    \\ \hline
\begin{tabular}[c]{@{}c@{}} \textbf{Augmented} \\ \textbf{Reality} \end{tabular}                                                                  &  Both                                                                                    & $10$                                                                                        & \begin{tabular}[c]{@{}c@{}}Variable\\ $[64, 1500]$\end{tabular}                     \\ \hline
\begin{tabular}[c]{@{}c@{}} \textbf{Configuration}  \\ \textbf{and Diagnostic} \end{tabular}                                                       &  Sporadic                                                                                & $[10, 100]$                                                                    & \begin{tabular}[c]{@{}c@{}}Variable \\ $[500, 1500]$\end{tabular}                   \\ \hline
\textbf{Best Effort}                                                                                                          &  Sporadic                                                                                & N.A.                                                                                      & \begin{tabular}[c]{@{}c@{}}Variable\\ $[30, 1500]$ \end{tabular}                     \\ \hline
\end{tabular}
\end{table}

 \begin{figure*}[t!]
    \centering
    \includegraphics[width=0.85\textwidth]{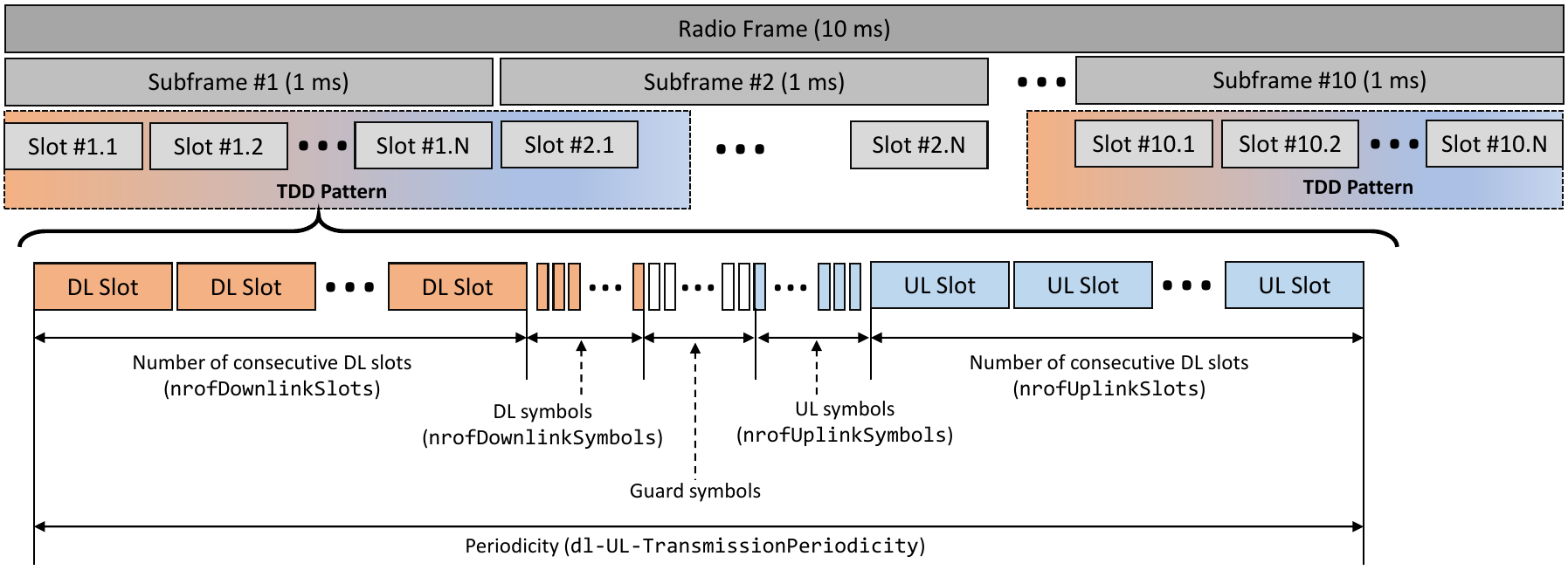}
    \caption{5G frame structure and TDD pattern.}
    \label{fig:TDDpattern}
\end{figure*}

\subsection{TDD Patterns}
The \gls{5G}-\gls{NR} frame structure is highly flexible, supporting various services with different requirements. As shown in Fig.~\ref{fig:TDDpattern}, it consists of a 10 ms frame divided into ten 1 ms subframes, each containing multiple 14-OFDM-symbol slots. The number of slots per subframe depends on the numerology parameter $\mu$, which defines the subcarrier spacing as $15 \cdot 2^{\mu} \, \text{KHz}$. Increasing $\mu$ reduces slot duration to $1/2^\mu$ milliseconds, enabling finer slot allocation for delay-sensitive services at the cost of higher bandwidth. The frame structure also supports mini-slots, short bursts of 2–6 \gls{OFDM} symbols, within a slot to further minimize latency.

The \gls{5G} frame structure's flexibility is further enhanced in \gls{TDD} mode, where \gls{DL} and \gls{UL} share the same frequency band at different times. \gls{TDD} patterns determine the slot allocation between \gls{UL} and \gls{DL}, optimizing data rate and latency based on application needs. A slot can contain \gls{DL}, \gls{UL}, or flexible \gls{OFDM} symbols, which can be configured for either transmission type. The \texttt{TDD-UL-DL-configurationCommon} parameter, defined in \gls{3GPP} TS 38.331 (v18.4.0), is key to allocate slots for \gls{DL} and \gls{UL} transmissions. Configured at the network level, it is communicated to each \gls{UE} via \gls{RRC} signaling. Fig.~\ref{fig:TDDpattern} shows the main elements of \texttt{TDD-UL-DL-configurationCommon}:

\begin{itemize}
    \item \textbf{Periodicity}: defines the time period over which the \gls{UL}/\gls{DL} pattern repeats, with the following possible values (in milliseconds): 0.5, 0.625, 1, 1.25, 2, 2.5, 5, and 10. 
    \item \textbf{Number of Consecutive \gls{DL} Slots}: specifies the number of consecutive full \gls{DL} slots at the beginning of each \gls{DL}/\gls{UL} pattern.
    \item \textbf{Number of Adjacent \gls{DL} Symbols}: defines the number of consecutive \gls{DL} symbols at the start of the slot following the last full \gls{DL} slot.
    \item \textbf{Number of Adjacent \gls{UL} Symbols}: refers to the number of consecutive \gls{UL} symbols at the end of the slot, preceding the first full \gls{UL} slot.
    \item \textbf{Number of Consecutive \gls{UL} Slots}: specifies the number of consecutive full \gls{UL} slots at the end of each \gls{DL}/\gls{UL} pattern.
    \item \textbf{Guard Period}: necessary for the transceiver to bridge from \gls{DL} to \gls{UL} and allow timing advance in \gls{UL}.
\end{itemize}

In this work, we evaluate different \gls{TDD} pattern configurations considering various values for the periodicity, as well as the allocation of slots between \gls{DL} and \gls{UL}.

\subsection{UE Buffer Status Report (BSR)}
The \gls{BSR} is a periodic \gls{MAC} layer message in \gls{5G} \gls{NR} sent from the \gls{UE} to the base station to report a quantized value of the buffered data awaiting transmission~\cite{3gpp_ts_38_321_v18_4_0}. This report, reflecting data in the \gls{RLC} and \gls{PDCP} layers, enables the \gls{UE} to request \gls{UL} grants, allowing the base station to allocate the necessary radio resources for data transmission.

In particular, a special case of the \gls{BSR} allows the \gls{UE} to explicitly indicate that its buffer is full. This mechanism, known as full \gls{BSR}, enables the base station to allocate \gls{UL} time slots more aggressively, assuming the \gls{UE} always has data to transmit. The full \gls{BSR} may be sent only once, when the \gls{UE} initially establishes communication with the base station, signaling that its buffer remains full. In our study, we analyze two \gls{BSR} transmission scenarios in a \gls{5G} \gls{TDD} system with a single \gls{UE}. In the first scenario, the \gls{UE} reports the actual data volume in its buffer, leading to dynamic \gls{UL} scheduling based on real-time buffer status. In the second scenario, the \gls{UE} sends a full \gls{BSR}, allowing the \gls{UE} to transmit user plane data as frequently as possible, minimizing latency without requiring further \gls{BSR} updates.

\section{Proof of Concept and Performance Results}\label{sec:ProofConceptandResults}
\subsection{Testbed Description}
\begin{figure}[b!]
    \centering
    \includegraphics[width=0.9\columnwidth]{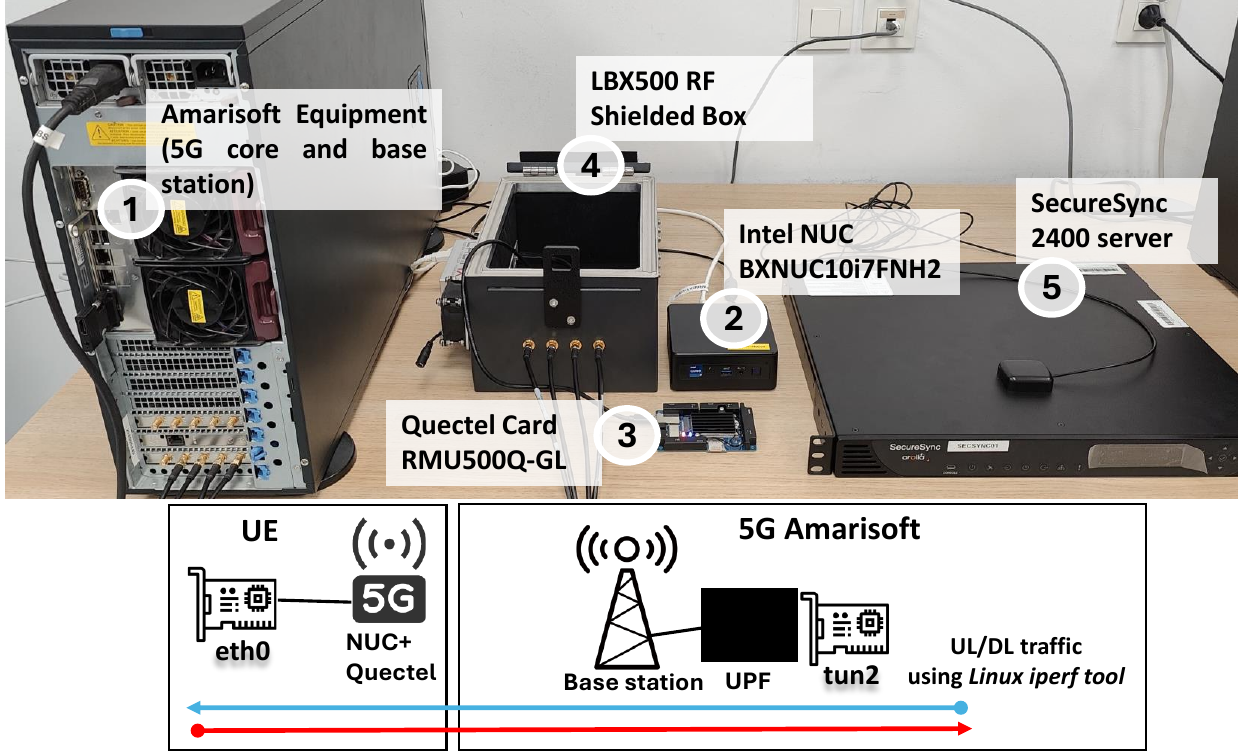}
    \caption{Proof of concept equipment and evaluated scenario.}
    \label{fig:testbed}
\end{figure}

To validate multiple \gls{TDD} pattern configurations, we implemented a testbed consisting solely of a commercial IP-based \gls{5G} system, as illustrated in Fig.~\ref{fig:testbed}. It comprises seven devices. A general-purpose computer equipped with a PCIe SDR50 card, referred to as \gls{5G} Amarisoft (Equip. 1), equipped with an Intel(R) Xeon(R) Bronze 3206R CPU @1.90GHz and 32GB RAM, runs the Amarisoft software to provide both the core and radio access network capabilities for a standalone \gls{5G} network. The testbed also includes one \gls{UE} consisting of an Intel NUC BXNUC10I7FNH2 (Equip. 2) paired with a Quectel RM500Q-GL card in an RMU500EK evaluation board (Equip. 3). This board uses the RM500QGLABR11A06M4G firmware and the NUC is equipped with an 11th Gen Intel(R) Core(TM) i7-1165G7 @2.80GHz and 16GB RAM. Both the \gls{5G} Amarisoft and \gls{UE} operate on Ubuntu 18.04.6 LTS. Since this system works in licensed bands, it is enclosed in an RF Shielded Test Enclosure, specifically the Labifix LBX500 model (Equip. 4). The last component is a SecureSync 2400 time synchronization server (Equip. 5), which distributes time using the \gls{NTP} to ensure time synchronization across devices.

\subsection{Experimental Setup}\label{sec:ExperimentalSetup}
We considered four \gls{TDD} patterns as summarized in Table~\ref{tab:tdd_patterns}: \gls{TDD} 44, \gls{TDD} 36, \gls{TDD} 72, and \gls{TDD} 10. The numbers following \gls{TDD} indicate the consecutive full slots allocated to \gls{DL} and \gls{UL}, respectively. All patterns have a periodicity of 10 slots between \gls{UL} and \gls{DL}, except for \gls{TDD} 10, which has a periodicity of 2 slots. For \gls{TDD} 10, although no dedicated \gls{UL} slot is provided, 12 out of 14 \gls{UL} \gls{OFDM} symbols are used for data transmission (i.e., as \texttt{nrofUplinkSymbols}), with the remaining symbols acting guard band. For all \gls{TDD} patterns, the slot duration is 0.5 ms, the minimum duration supported by the Quectel card (i.e, $\mu$=1), allowing for the lowest possible packet transmission delay with our equipment. The base station operates with a bandwidth of 50 MHz and transmits in band n79 (4,4 GHz - 5 GHz).
\begin{table}[b!]
    \centering
    \caption{Evaluated TDD Patterns}
    \resizebox{1.0\columnwidth}{!}{
    \begin{tabular}{|c|c|c|c|c|}
        \hline
        \textbf{TDD Pattern} & \textbf{DL Slots} & \textbf{UL Slots} & \textbf{Guard Band} & \textbf{Periodicity} \\
        \hline
        TDD 44 & 4 & 4 & 2 slots & 10 slots \\
        TDD 36 & 3 & 6 & 1 slot & 10 slots \\
        TDD 72 & 7 & 2 & 1 slot & 10 slots \\
        TDD 10 & 1 & 12 symbols & 2 symbols & 2 slots\\
        \hline
    \end{tabular}
    }
    \label{tab:tdd_patterns}
\end{table}

We aim to measure the packet transmission delay in the \gls{5G} system for both \gls{UL} and \gls{DL} directions, using the outlined \gls{TDD} patterns. The measurement points within the \gls{5G} system are illustrated in the bottom image of Fig.~\ref{fig:testbed}. Specifically, these points are the \gls{NIC} of the \gls{5G} system, via the \gls{UPF}, and the \gls{NIC} of the \gls{UE}. To measure the delay, we used \texttt{tcpdump} to capture traces from the \glspl{NIC}, identifying packets via the IP header's ID field. Timestamps were extracted and subtracted to calculate the delay. To emulate periodic industrial traffic, we used the Linux tool \texttt{iperf} in \texttt{UDP} mode, with the server and client configured based on traffic direction. Additionally, our work includes measurements of throughput, which refers to the data rate generated by the transmitter, and effective throughput, which represents the rate of data successfully received by the receiver. 

For each \gls{TDD} pattern, we considered all possible combinations of the following parameters: (a) four traffic load levels, where the same load was applied simultaneously to both \gls{UL} and \gls{DL} directions at 10 Mbps, 20 Mbps, 30 Mbps, and 40 Mbps; (b) two packet sizes, using either 100 bytes or 1000 bytes; and (c) two configurations for the \gls{BSR}, one where the full \gls{BSR} is active and another where the full \gls{BSR} is not active. A total of 100,000 packets was generated in each direction for all possible scenarios to provide a sufficient number of samples to derive the distribution of the packet transmission delay.

The dataset is made available to foster reproducibility\footnote{Online Available: \url{https://github.com/wimunet/empirical_analysis_5G_tdd_patterns_for_industry_4_0}}.

\subsection{Performance Results}
 \begin{figure*}[t!]
    \centering
    \includegraphics[width=0.86\textwidth]{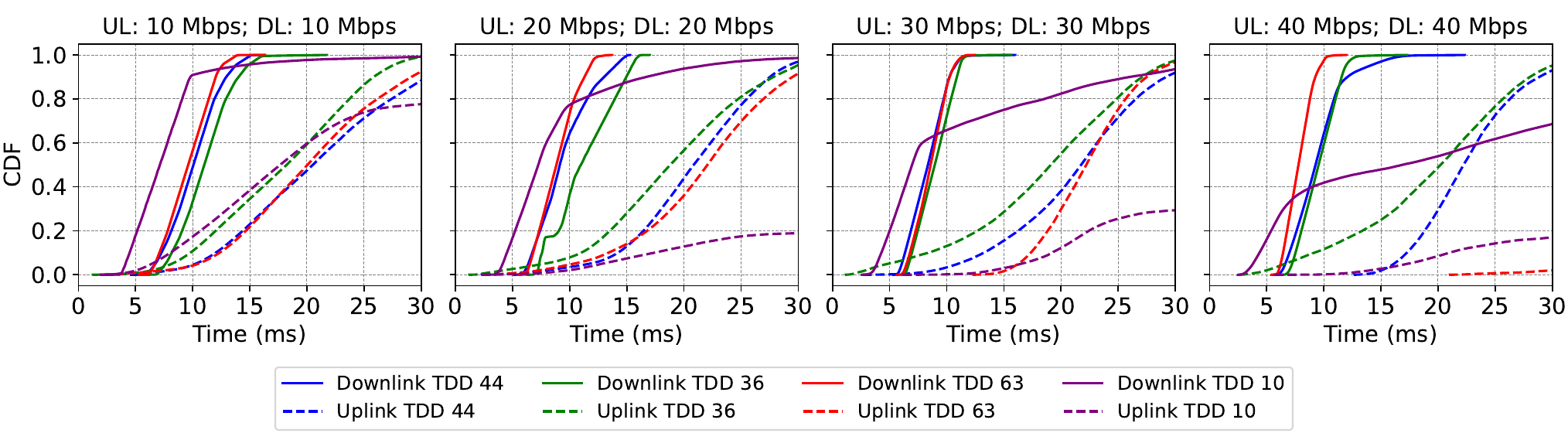}
    \caption{CDF of the packet transmission delay for DL and UL with different TDD patterns when full \gls{BSR} is not activated. The packet size is 100 bytes.}
    \label{fig:results_100bytesnoBSR}
\end{figure*}

\begin{figure*}[t!]
    \centering
    \includegraphics[width=0.86\textwidth]{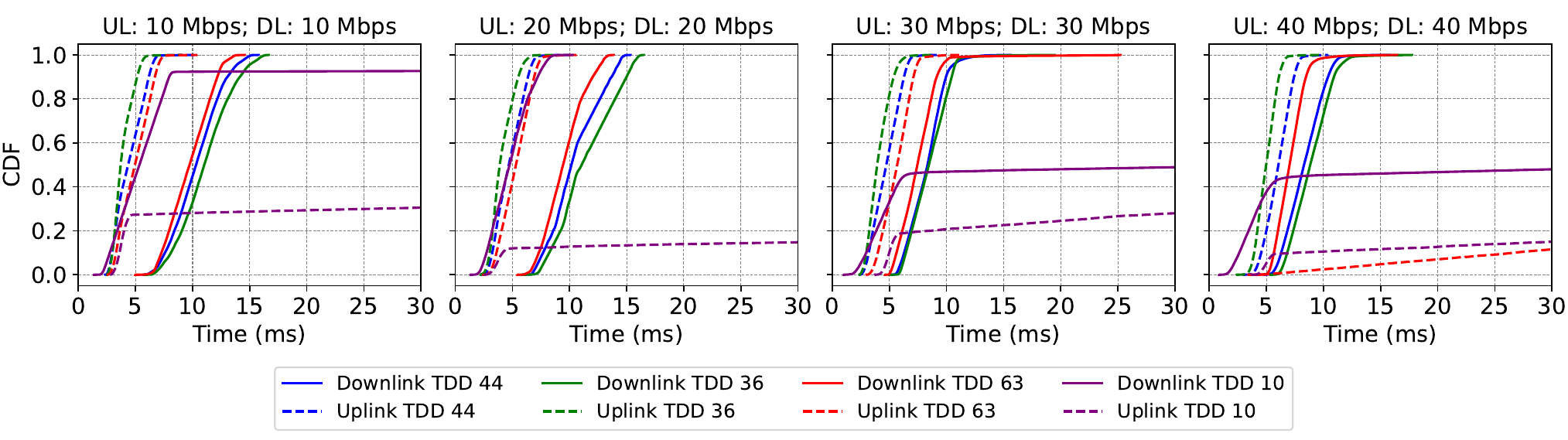}
    \caption{CDF of the packet transmission delay for DL and UL with different TDD patterns when full \gls{BSR} is activated. The packet size is 100 bytes.}
    \label{fig:results_100bytesBSR}
\end{figure*}

\begin{figure*}[t!]
    \centering
    \includegraphics[width=0.86\textwidth]{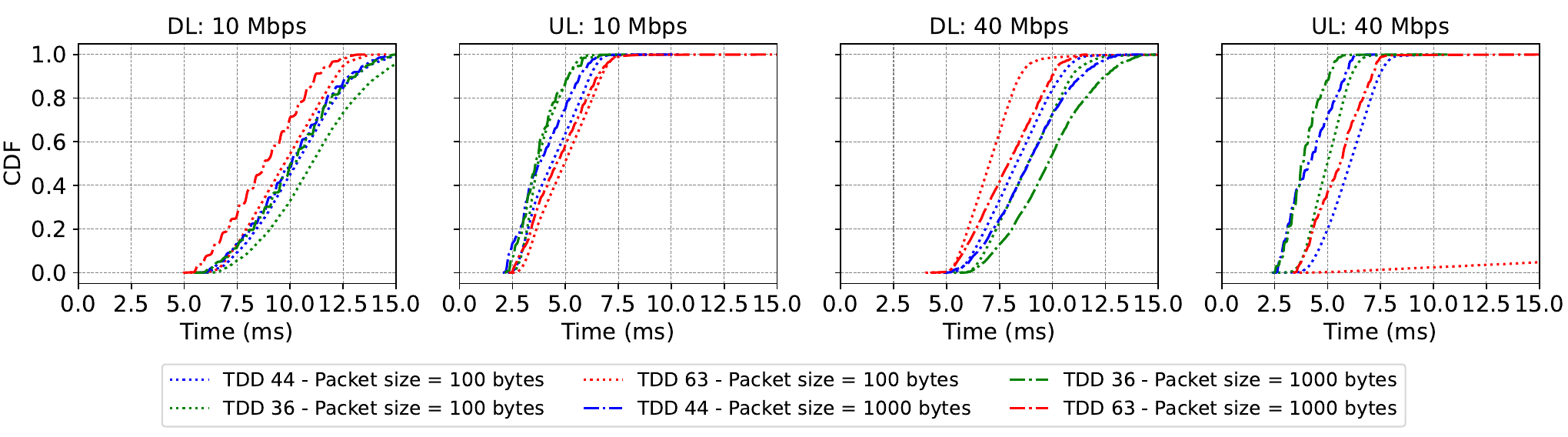}
    \caption{CDF of the packet transmission delay for different TDD patterns, comparing 100-bytes and 1000-bytes packet sizes. Full BSR is activated.}
    \label{fig:100vs1000}
\end{figure*}

Fig.~\ref{fig:results_100bytesnoBSR} shows the \gls{CDF} of packet transmission delay for 100-byte packets with full \gls{BSR} disabled, under varying traffic loads. First, without full \gls{BSR}, \gls{DL} delay is significantly lower than \gls{UL}. It is due to the \gls{UE} periodically sending \gls{BSR} to request resources, thereby wasting transmission slots in the control plane. Second, among different \gls{TDD} patterns, \gls{TDD} 36 (green curves) has the highest \gls{DL} delay and the lowest \gls{UL} delay, due to allocating more timeslots for \gls{UL}. In contrast, \gls{TDD} 63 (red curves) shows the opposite effect, with the highest \gls{UL} delay and the lowest \gls{DL} delay. \gls{TDD} 44 (blue curves) exhibits intermediate behavior. \gls{TDD} 10 (purple curves), with a reduced repetition period, results in much lower delays for both \gls{DL} and \gls{UL} at 10 Mbps, but limits base station capacity. For \gls{DL} traffic at 10, 20, 30, and 40 Mbps, the effective throughput is 97.1\%, 76.5\%, 65.3\%, and 64.0\%, respectively. In \gls{UL}, throughput degradation is sharper, with 98.7\%, 90.5\%, 30.7\%, and 3.93\% effective throughput, respectively. This is mainly due to the signaling needed for \gls{UE} to send \gls{BSR} messages and request transmission grants. As the traffic load increases, the base station takes longer to respond to these grants, exacerbating the delay. Finally, while \gls{TDD} 44, \gls{TDD} 36 and \gls{TDD} 63 patterns show increased delays with higher traffic, they generally maintain acceptable latency for industrial automation applications in \gls{DL}. However, \gls{UL} latency becomes problematic, especially for \gls{TDD} 36, as 10\% of the packets experience a transmission delay exceeding 30 ms at 40 Mbps, which represents an excessively high ratio for some industrial applications.

Fig.~\ref{fig:results_100bytesBSR} shows the \gls{CDF} of packet transmission delay for 100-byte packets with the full \gls{BSR} mechanism enabled. Notably, full \gls{BSR} significantly reduces \gls{UL} delay compared to \gls{DL}, as the \gls{UE} no longer needs to report buffer status continuously. Instead, the \gls{UE} signals a full transmission buffer to the base station, which then proactively allocates \gls{UL} slots for data transmission, eliminating the need for periodic reporting. Furthermore, \gls{DL} delay remains unaffected by the activation of full \gls{BSR}. A key drawback occurs with reduced \gls{TDD} pattern periodicity, such as \gls{TDD} 10. With more \gls{UL} slots allocated for user plane data, the \gls{5G} system reaches its effective throughput limit with lower traffic loads than without full \gls{BSR}, resulting in increased \gls{UL} delays at high traffic loads.

Fig.~\ref{fig:100vs1000} presents the \gls{CDF} of the packet transmission delay considering different packet sizes under low (i.e., 10 Mbps) and high (i.e., 40 Mbps) traffic load scenarios. In this figure, the dotted curves represent the transmission of 100-byte packets, while the dash-dotted curves represent the transmission of 1000-byte packets. We observe that, for all \gls{TDD} patterns, the packet transmission delay is consistently lower when using 1000-byte packets in all scenarios, except for \gls{DL} under high traffic load. The lower packet transmission delay observed in most cases when using 1000-byte packets can be explained by the reduced number of Layer 3 packets that the receiver (either the \gls{UE} or the base station) needs to reconstruct from the \gls{5G} transport blocks\footnote{A 5G transport block is a data unit transmitted at the physical layer, while a packet is a Layer-3 data unit which includes routing information.}. Since the data rate remains constant, using packets of greater size results in fewer packets to reconstruct. However, in \gls{DL} with high traffic load, the packet transmission delay is higher when using 1000-byte packets. This is because the \gls{UE} requires more processing time to construct individual packets from the transport blocks when the packets are larger in size. This behavior is explained by the fact that the processing capacity of the Quectel card is significantly lower than that of the PC hardware which runs the Amarisoft base station. It is reasonable to observe that a \gls{UE} has lower processing capabilities compared to the base station's hardware. Finally, in most cases, the time difference between transmitting 1000-byte and 100-byte packets ranges from approximately 0.5 ms to 1.5 ms, which can be significant for isochronous and cyclic synchronous traffic.

\subsection{Recommendations for 5G TDD Pattern Configurations}\label{sec:Recommendations}

After analyzing our results, we aim to provide a series of configuration recommendations for \gls{5G} \gls{TDD} patterns that will better support industrial automation traffic. They are:
\begin{itemize}
    \item[R1] \textbf{Ensure Deterministic Communication for Industrial Traffic}: For industrial automation, it’s crucial that packet delays remain below the threshold with a very high probability (e.g., 99.999\%), ensuring reliable and timely communication. Monitor the tail of the CDF is key to confirm delays stay within the defined bound.
    \item[R2] \textbf{Perform Traffic Load Assessment for \gls{UL}/\gls{DL}}: It is advisable to conduct a prior study on the traffic load expected for \gls{UL}/\gls{DL} in the considered industrial environment. This will allow for proper allocation of time slots between \gls{UL} and \gls{DL}, ensuring that both directions have enough resources for efficient data transmission.
    \item[R3] \textbf{Activate Full \gls{BSR} for Latency-Critical UL Traffic}: In scenarios where \gls{UL} traffic is not critical in terms of latency, activating full \gls{BSR} is not necessary. Otherwise, activating this parameter is highly recommended because it significantly reduces the signaling sent from the \gls{UE} in \gls{UL}, which, in turn, optimizes the available time slots for transmitting user plane data and ultimately leads to a latency reduction.
    \item[R4] \textbf{Consider Traffic Load for \gls{TDD} Pattern Periodicity}: While it may seem intuitive that reducing the periodicity of \gls{TDD} patterns would decrease packet transmission delay, this approach only works effectively when the \gls{5G} system's traffic load is low. In scenarios with a high traffic load, packet transmission delays can increase significantly, making this configuration unsuitable for many types of industrial automation traffic.
    \item[R5] \textbf{Consider Packet Size for Traffic with very Low Latency Requirements}: Considering the packet size of specific industrial automation traffic is key, especially when this traffic has stringent delay requirements (e.g., packet delay budget of around 2 ms or less). The packet size can significantly affect transmission delay, and optimizing this parameter is crucial to meeting the latency requirements of time-sensitive applications, particularly for isochronous and cyclic synchronous traffic.
\end{itemize}

\section{Conclusions and Future Works}\label{sec:conclusions}
In \gls{5G}-\gls{TSN}-based industrial networks, ensuring deterministic \gls{E2E} packet transmission delay is vital for industrial automation, requiring delays under a predefined threshold, e.g., 20 ms with 99.999\% reliability. Achieving this determinism is challenging, especially in \gls{5G} systems, due to the variable delays introduced by wireless transmission. The \gls{5G} \gls{TDD} pattern configuration, which allocates transmission slots between \gls{UL} and \gls{DL}, directly affects the packet transmission delay. Most research on \gls{5G} \gls{TDD} configurations relies on simulations, neglecting real-world industrial challenges, especially the deterministic delay and reliability requirements critical for industrial automation. This paper investigates the impact of different \gls{TDD} patterns on \gls{5G} systems based on empirical measurements taken from a base station in a laboratory testbed. We assess packet transmission delay under varying conditions such as traffic load, packet size, and the effect of full \gls{BSR} activation at the \gls{UE}. Our results yield key recommendations: \textit{i}) Ensure deterministic communication by monitoring the CDF tail to keep delays within bounds; \textit{ii}) Assessing \gls{UL} and \gls{DL} traffic load helps allocate time slots efficiently; \textit{iii}) Activating full \gls{BSR} reduces latency for \gls{UL} traffic by minimizing signaling overhead; \textit{iv}) Reducing \gls{TDD} pattern periodicity can increase delays under high load; \textit{v}) Packet size plays a crucial role in meeting the stringent latency requirements of highly time-sensitive applications.

Future work will explore additional \gls{TDD} pattern configurations on our testbed, including scenarios with multiple \glspl{UE}, \gls{5G} \gls{QoS} flows, network slicing, multiple cell deployments and \gls{TSN} bridges.

\section*{Acknowledgment}
This work has been financially supported by the Ministry for Digital Transformation and of Civil Service of the Spanish Government through TSI-063000-2021-28 (6G-CHRONOS) project, and by the European Union through the Recovery, Transformation and Resilience Plan - NextGenerationEU. Additionally, this publication is part of grant PID2022-137329OB-C43 funded by MICIU/AEI/ 10.13039/501100011033 and by ERDF/EU. Finally, this publication is also part of grant FPU20/02621 funded by the Spanish Ministry of Universities.

\bibliographystyle{ieeetr}
\bibliography{references}

\end{document}